\documentstyle[epsfig]{aipproc}
\begin{document}

\title{Signal Characteristics from Electromagnetic Cascades in Ice}
\author{Soebur Razzaque, Surujhdeo Seunarine, David Z. Besson, and
Douglas W. McKay}
\address{Department of Physics and Astronomy \\ 
University of Kansas, Lawrence, Kansas 66045-2151}
\maketitle

\begin{abstract}
We investigate the development of electromagnetic cascades in ice
using a GEANT Monte Carlo simulation. We examine the Cherenkov pulse
that is generated by the charge excess that develops and propagates
with the shower. This study is important for the RICE experiment at
the South Pole, as well as any test beam experiment which seeks to
measure coherent Cherenkov radiation from an electromagnetic shower.
\end{abstract}

\section{Introduction}
Ultra high energy neutrinos can travel a long distance without
scattering. By detecting the sources of these particles, we can probe
the physics of the standard model (and beyond) at energies
un-obtainable at current accelerators. An ultra high energy neutrino
can interact via a charge current interaction, giving most of its
energy to the secondary electron, which can then initiate an
electromagnetic cascade or shower. Askaryan \cite{askaryan62}
predicted a negative charge imbalance in the cascade which gives rise
to coherent Cherenkov radiation at radio frequencies. The predicted
flux of ultra high energy neutrinos is small and model dependent
\cite{Stecker96,halzen97,rachen99}. An experiment to detect ultra high
energy neutrinos using radio antennas requires a large volume because
of the small flux. A dense, radio-transparent target is needed for the
small shower size needed for coherence and small signal
attenuaton. Antarctic ice is suitable for this purpose. A detailed
analysis of such an experiment was done \cite{fmr96} which concluded
that a modest array of antennas can detect many events per year. The
Radio Ice Cherenkov Experiment (RICE) at the South Pole \cite{rice99}
is a prototype designed to detect neutrinos with energy $\ge\;PeV$
using this method.  A reliable Monte Carlo simulation tool is needed
to study the shower development, Cherenkov radiation, detector, and
data acquisiton system.  One can also test the idea of coherent
Cherenkov emission at accelerator facilities by dumping bunches of
electrons or photons in a dense target like sand or salt or any other
suitable medium \cite{gorham00}. Zas, Halzen and Stanev (ZHS)
\cite{zhs92} developed a Monte Carlo simulation to study
electromagnetic showers and Cherenkov emission in ice. Buniy and
Ralston \cite{roman00} have also developed a method to estimate the
coherent Cherenkov signal by parametrizing the cascades.

We have developed a GEANT Monte Carlo simulation primarily to study
the coherent Cherenkov emmission in ice but it can easily be modified
for other similar studies with different materials.  GEANT
\cite{geantman} is a well known and widely used detector simulation
package in particle physics.  It allows access to all details of the
simulation such as controls of various processes, definition of target
and detector media, and a complete history of all events simulated. It
can be used to simulate all dominant processes in the
$10\;keV-10\;TeV$ energy range.  Cross sections of electromagnetic
processes are reproduced in GEANT within a few percent up to a hundred
$GeV$. We use GEANT to simulate electromagnetic cascades inside
materials from which we extract track information. We take this track
information and determine the resulting radio pulse using standard
electrodynamics calculations. We also calculate other shower
parameters using the track and energy information from GEANT.
\section{Shower Description}
When a high energy electron ($e^-$) or photon ($\gamma$) hits a
material target, an electromagnetic cascade is created inside the
material.  {\it Bremmstrahlung} and {\it pair production} are the
dominant high energy processes at the beginning of the shower
development.  Due to bremmstrahlung, an $e^-$ loses $1/e$ of it's
energy on average over a distance $X_0$, the {\it radiation
length}. The secondary $\gamma$ can then produce an $e^+\;e^-$
pair. The number of particles thus grows exponentially. $e^-$ and
$e^+$ lose energy due to {\it ionization} as they travel inside the
material. After reaching a {\it critical energy}($E_c$), when the
energy loss due to bremmstrahlung is equal the energy loss due to
ionization, $e^-$, $e^+$ lose their energy mostly due to ionization
and the cascade eventually stops. A rough estimate of the critical
energy is $E_c \approx 605/Z \; MeV$ which can be obtained by equating
the radiation and the ionization loss formulae. $Z$ is the atomic
number of the medium.

All models of an electromagnetic cascade, including a simplified model
developed by Heitler \cite{heitler}, and a more realistic model
developed by Carlson and Oppenheimer \cite{oppenheimer37}, show
similar basic features which include linear scaling of the track
lengths and logarithmic scaling of the maximum position with primary
energy.

The cascade is concentrated near an axis along the direction of the
primary.  {\it Multiple Couloumb scattering} is responsible for the
transverse spread of the cascade. A quantity called {\it Moliere
radius} ($R_M$) is determined by the average angular deflection per
radiation length at the critical energy ($E_c$) and is used to
estimate the transverse spread. One can also look at the fraction of
energy that escapes transverse to the shower axis
\cite{nelson66,bathow70} to get a good estimate of $R_M$. About $90\%$
of the primary energy is contained inside a tube of radius $R_M$ along
the shower axis. A relationship between the Moliere radius, the
critical energy and the radiation length is $R_M=X_0 E_s/E_c$ where
$E_s\approx 21.2\;MeV$ is the scale energy.

Low energy processes like {\it Compton}, {\it Moller} and {\it Bhabha}
scatterings and {\it positron annihilation} build up a net charge
(more $e^-$ than $e^+$) in the cascade, as atomic electrons in the
target medium are swept up into the forward moving shower.
\section{Electromagnetic Pulse}
\begin{figure}[b!] 
\centerline{\epsfig{file=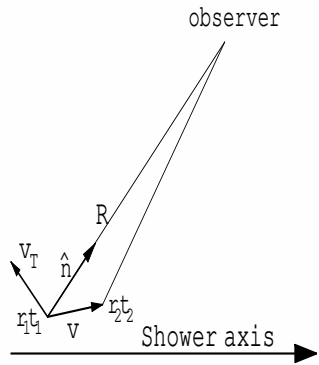,height=2.8in,width=2.8in}}
\caption{Set-up to calculate electromagnetic pulse from a single track.}
\label{fig:diag}
\end{figure}

High energy $e^-$ and $e^+$ in the cascade can travel faster than the
speed of light in the medium and give rise to Cherenkov radiation. The
electric field due to a single charge moving uniformly from position
${\vec r}_1$ to ${\vec r}_2={\vec r}_1+{\vec v}\delta t$ in the {\it
Fraunhoffer} limit, is given by
\begin{equation}
R{\vec E}(\omega)=\frac{\mu_r}{\sqrt{2\pi}}\left(\frac{e}{c^2}\right)
e^{i\omega \frac{R}{c}}
e^{i\omega(t_1-n{\vec \beta}\cdot{\vec r}_1)} {\vec v}_T 
\frac{(e^{i\omega \delta t
(1-{\hat n}\cdot{\vec \beta}n)}-1)}{1-{\hat n}\cdot{\vec \beta}n} 
\label{eq:pulse2}
\end{equation}
where $\mu_r$ is the relative permeability and $n$ is the refractive
index of the medium. The condition $1-{\hat n}\cdot{\vec \beta}n=0$
defines the Cherenkov angle $\theta_c$ as $\cos\theta_c=1/n\beta$.  At
or very close to the Cherenkov angle or at low frequency, the equation
(\ref{eq:pulse2}) can be reduced to the form
\begin{equation}
R{\vec E}(\omega)=\frac{\mu_r i \omega}{\sqrt{2\pi}}\left(\frac{e}{c^2}\right)
e^{i\omega \frac{R}{c}}
e^{i\omega(t_1-n{\vec \beta}\cdot{\vec r}_1)} {\vec v}_T \delta t .
\label{eq:pulse3}
\end{equation}
See Fig. \ref{fig:diag} for a description of various quantities in
equations (\ref{eq:pulse2}, \ref{eq:pulse3}).
\section{Monte Carlo Results}
Since on average, the electrons lose $1/e$ of their energy in each
radiation length due to bremmstrahlung, one can fit an exponential to
the energy loss data generated by the Monte Carlo and find a value for
the radiation length ($X_0$).  This will serve as an internal
consistency check to ensure that we are tracking all the particles
along with their energies.  We generated $500$ showers of $50\;GeV$
electrons and recorded the energy loss due to bremsstrahlung.  A fit
to this gives a radiation length of $41.5 \pm 3.2 \;cm$. Given the
molecular composition, GEANT also calculates the medium's radiation
length.  For ice it is $38.8\;cm$, which is in agreement with the
value we calculated.

\begin{center}
\begin{tabular}{l|ccc}
\multicolumn{4}{c}{Table I GEANT consistency checks}\\
\hline \hline
Parameters     & Iron    & Lead   & Ice \\ \hline
$X_0$ ($cm$) (from ref \cite{pdg96})   & $1.76$  & $0.56$ & $39.05$ \\
$R_M$ ($cm$)   & $2.1$   & $1.6$  & $13$ \\
$E_c=X_0 E_s/R_M$ ($MeV$) & $17.77$ & $7.42$ & $63.7$ \\
$E_c=605/Z$ ($MeV$) & $23.3$  & $7.4$  & $83.8$ \\
\hline \hline
\end{tabular}
\end{center}

To calculate the Moliere radius ($R_M$) we construct an imaginary
cylinder centered on the shower axis (up to the physical length of the
shower). We add the energy ($U$), of all the tracks that leave the
cylinder without re-entering. By varying the radius of the cylinder we
arrive at $R_M$ which is the radius of the cylinder when the fraction
$U/E$ is equal to $0.1$, $E$ being the initial energy of the
cascade. We have checked $R_M$ for lead, iron and ice (see table
I). One can also calculate the critical energy ($E_c$) using two
different formulae quoted earlier. 
\begin{figure}[b!] 
\centerline{\epsfig{file=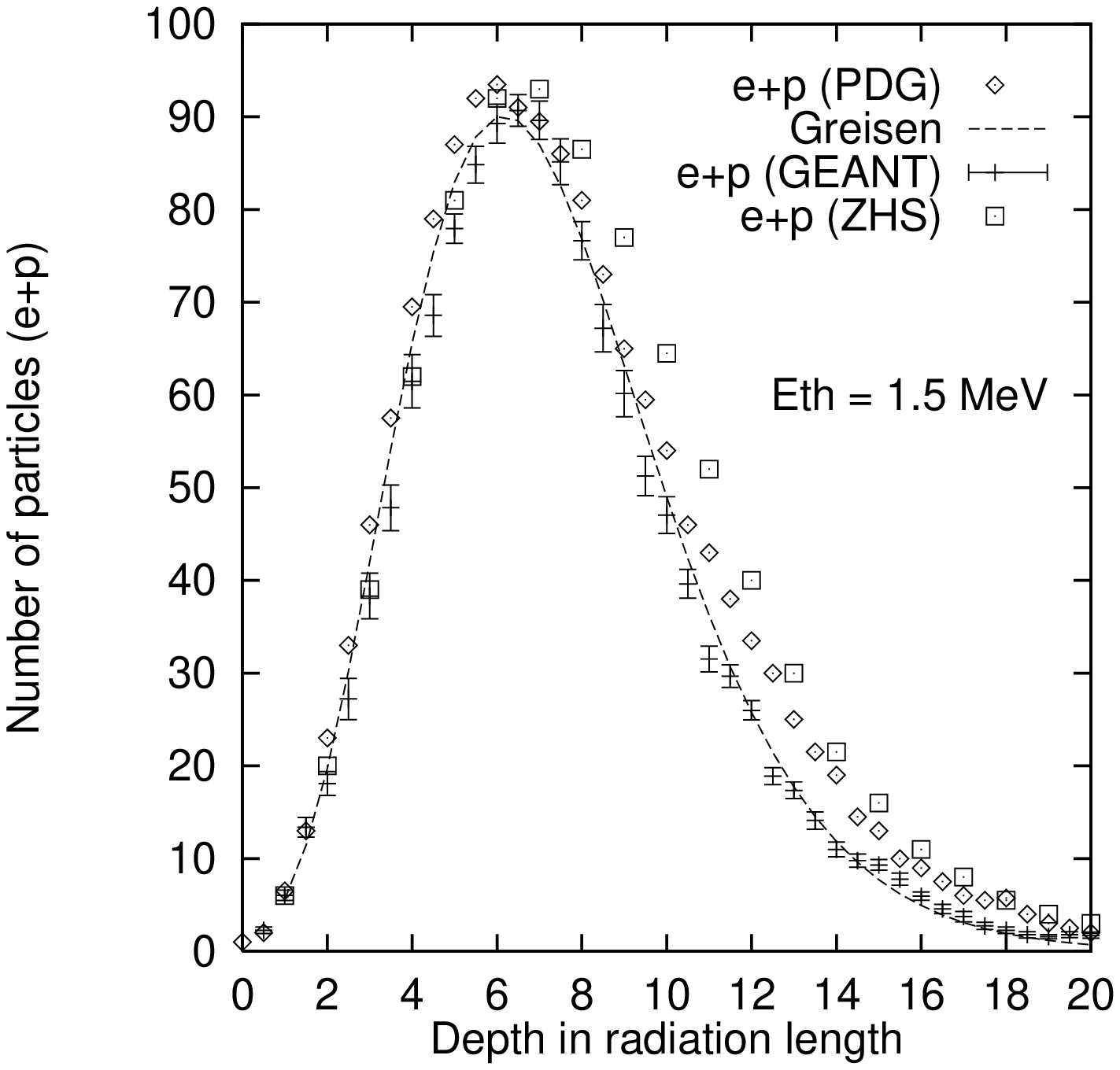,height=2.8in,width=2.8in} 
\epsfig{file=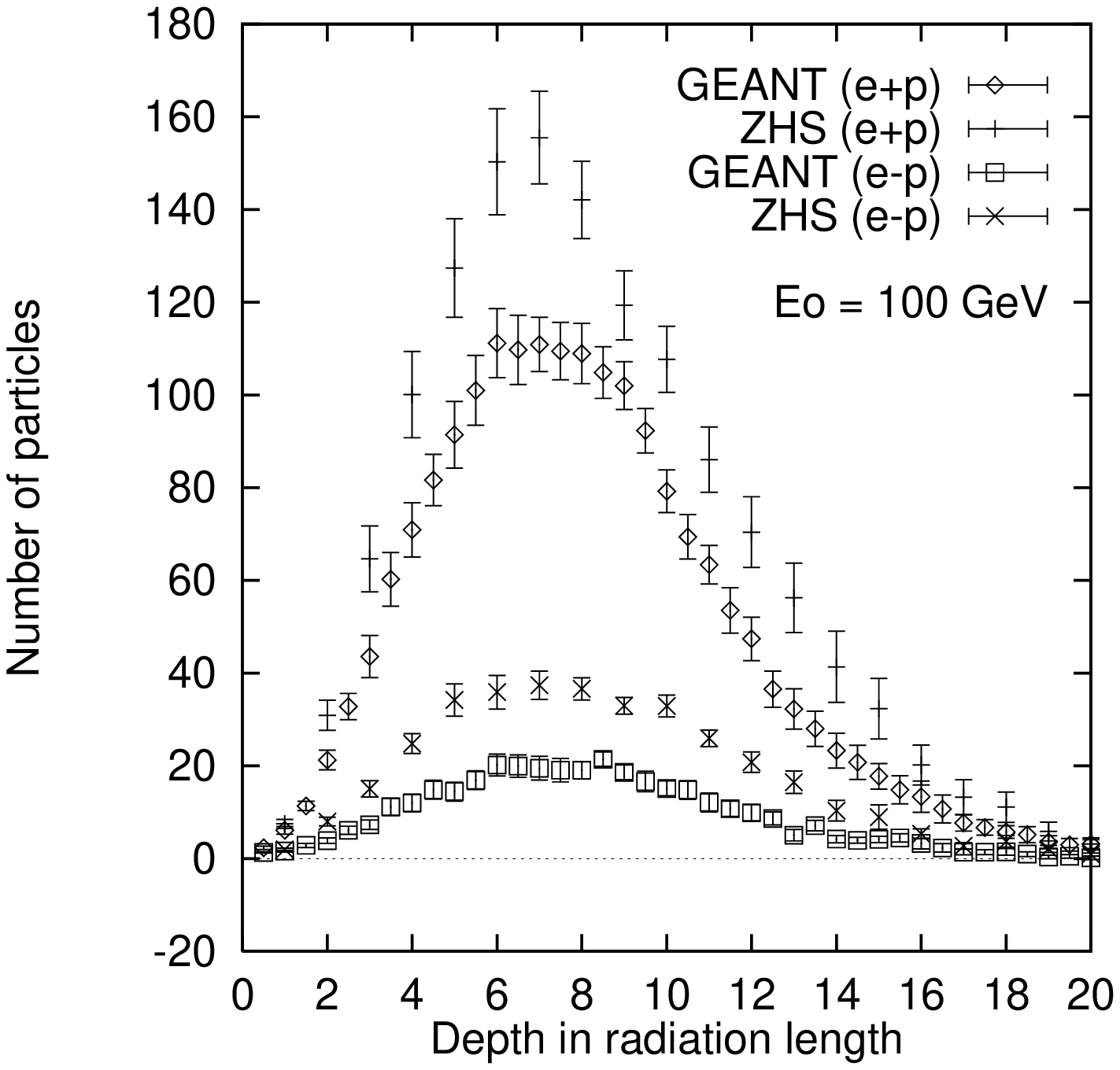,height=2.8in,width=2.8in}}
\vspace{10pt}
\caption{(a) The longitudinal profile of a 30 $GeV$ cascade in iron
using GEANT (average of 30 showers) and comparison with the same
produced by EGS4 (from particle data book) and by ZHS Monte
Carlos. The error bars correspond to standard error. (b) Comparisons
between the profiles of a 100 $GeV$ cascade (averaged over 20 showers
each) from GEANT and from ZHS Monte Carlos. The total energy threshold
is 0.611 $MeV$ in both cases.}
\label{fig:iron}
\end{figure}

Shower depth ($t$) is measured in units of radiation length ($X_0$) as
$t = x/X_0$. We have simulated $30\;GeV$ electron-induced cascades in
iron.  The longitudinal profile was obtained by adding the number of
particles with total energy greater than $1.5\;MeV$ crossing planes at
every half radiation length perpendicular to the shower axis as shown
in Fig. \ref{fig:iron}a. The number of particles agrees reasonably with
EGS4 simulation of the same shower \cite{pdg96}.  The Greisen-Rossi
distribution for total number of particles (electrons and positrons)
is shown by the dashed line. A least squares fit to the GEANT data
with Greisen-Rossi distribution gives about $90\%$ confidence level.

We used the same method to study the longitudinal profile of showers
in ice. Simulation shows the correct scaling behavior of the number of
particles with the initial energy of the shower. The position of the
shower maximum also shows the correct logarithmic scaling with the
initial energy. The charge excess ($\Delta Q$) is defined as $\Delta Q
= \frac{N(e)-N(p)}{N(e)+N(p)}$ where $N(e)$ and $N(p)$ are the number
of electrons and positrons respectively as functions of shower
depth. According to our simulations, the net charge imbalance is
$15\%-18\%$ at the shower max. Note that there is no direct
experimental data on this.

A comparison of $100\;GeV$ shower (averaged over 20 showers) with
$611\;KeV$ threshold from GEANT to the same from ZHS Monte Carlo
(Fig. \ref{fig:iron}b) shows about a $25\%-35\%$ discrepancy for the
total number of particles at the shower max. The discrepancy between
the two simulations remains the same at higher energies.

\begin{figure}[b!] 
\centerline{\epsfig{file=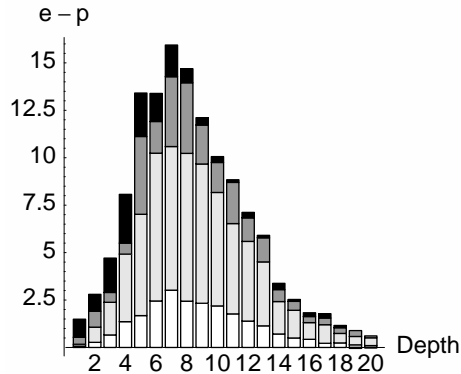,height=3.in,width=3.in}}
\caption{Excess charge plot for a 500 $GeV$ cascade (averaged over 20
showers) in ice using GEANT. The excess charge is shown in energy bins
(from bottom to top) 1.5-5 $MeV$ (white), 5-30 $MeV$ (light gray),
30-100 $MeV$ (dark gray) and above 100 $MeV$ (black).}
\label{fig:depth}
\end{figure}

In Fig. \ref{fig:depth}, we show the charge excess in a $500\;GeV$
shower from GEANT with kinetic energy threshold $1.5\;MeV$.  This is
an average of $20$ showers and the charge excess is broken down to
different energy bins. It shows that most of the contributions to
charge excess come from the energy range $1.5$ to $30\;MeV$.
The Cherenkov pulse from a cascade is proportional to the total track
length (energy deposition) of the cascade. Our results show the
correct scaling behavior of the track lengths with the initial energy
($E_o$). If we consider all processes to be elastic except ionization,
then an upper bound for the total track length is $L
=E_o/\left(\frac{dE}{dx}\right)_{ion}^{avg}$ where
$(\frac{dE}{dx})_{ion}^{avg}$ is the average ionization energy loss
per unit length. To determine ($\frac{dE}{dx}$) we generated $500$ of
$1\;GeV$ tracks and kept a record of the rate at which energy was lost
due to ionization. Fig.\ref{fig:pul}a shows the average
($\frac{dE}{dx}$) for the $500$ tracks. It can be seen that the
average value matches theoretical Bethe-Bloch result reasonably
well. The average ionization loss in the relativistic rise region is
approximately $1.9\;MeV/cm$.
\section{Pulse Calculation}
To calculate the electric pulse from the cascade, we added up the
contributions (\ref{eq:pulse2}) from all charged tracks. There are two
assumptions we made to evaluate pulse (\ref{eq:pulse2}) and
(\ref{eq:pulse3}) from GEANT track information. First, we assume an
azimuthal symmetry about the shower axis, which allows us to evaluate
pulse equations in 2-dimensions. This is a good approximation as long
as we have many tracks in the shower. A typical $500\;GeV$ shower has
thousands of tracks and the number goes up as we increase the
energy. We checked this approximation on a shower-by-shower basis and
did not find any noticeable difference between the pulses evaluated in
the $x-z$ plane and in the $y-z$ plane ($z$ - is the shower
axis). Second, we evaluate the times $t_1$ and $\delta t$ in
(\ref{eq:pulse2}) and (\ref{eq:pulse3}) assuming that the particles
travel at the speed of light. This is a good approximation if the
frequency is not too high as can be seen from (\ref{eq:pulse2}).
\begin{figure}[b!] 
\centerline{\epsfig{file=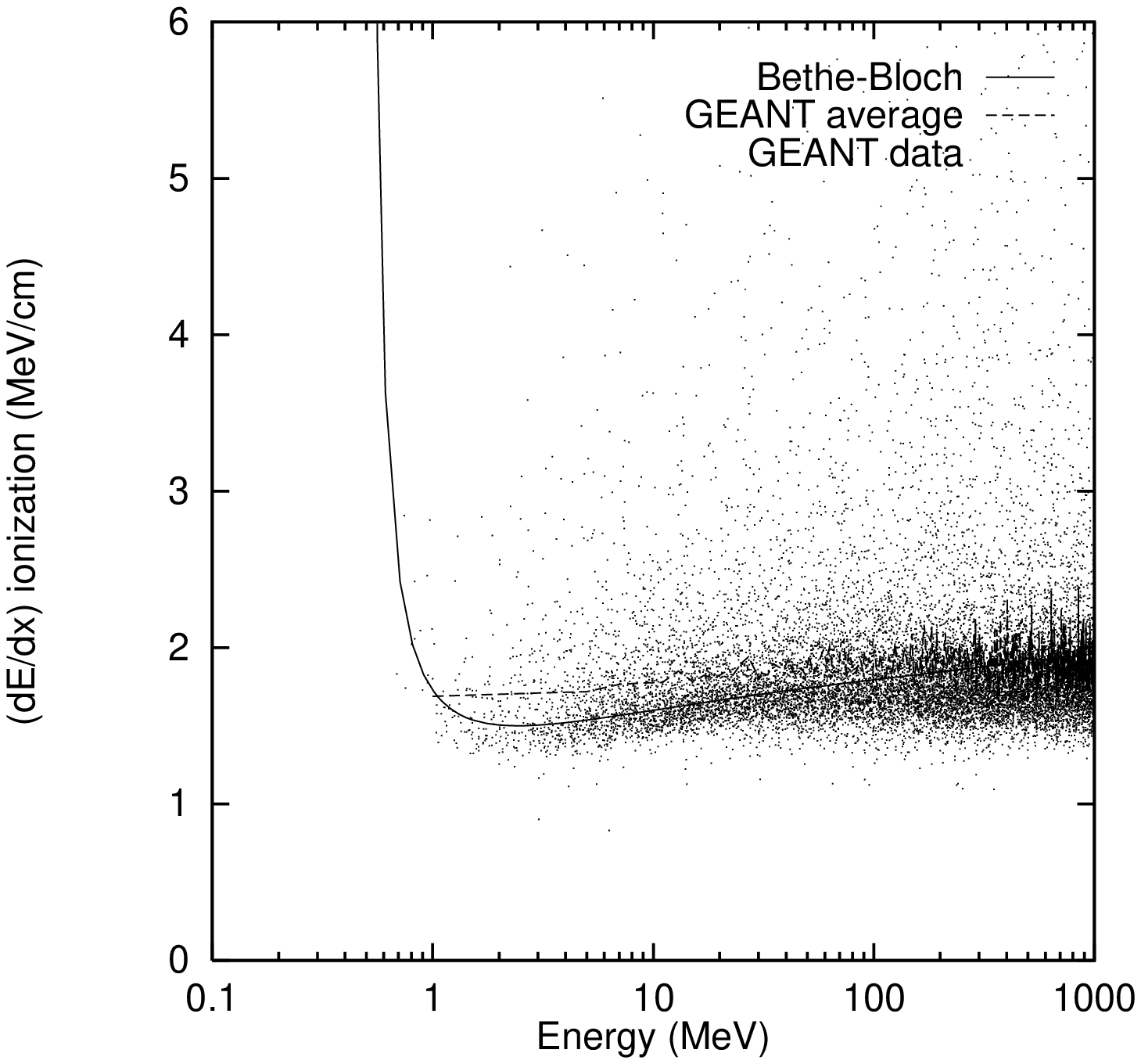,height=2.8in,width=2.8in}
\epsfig{file=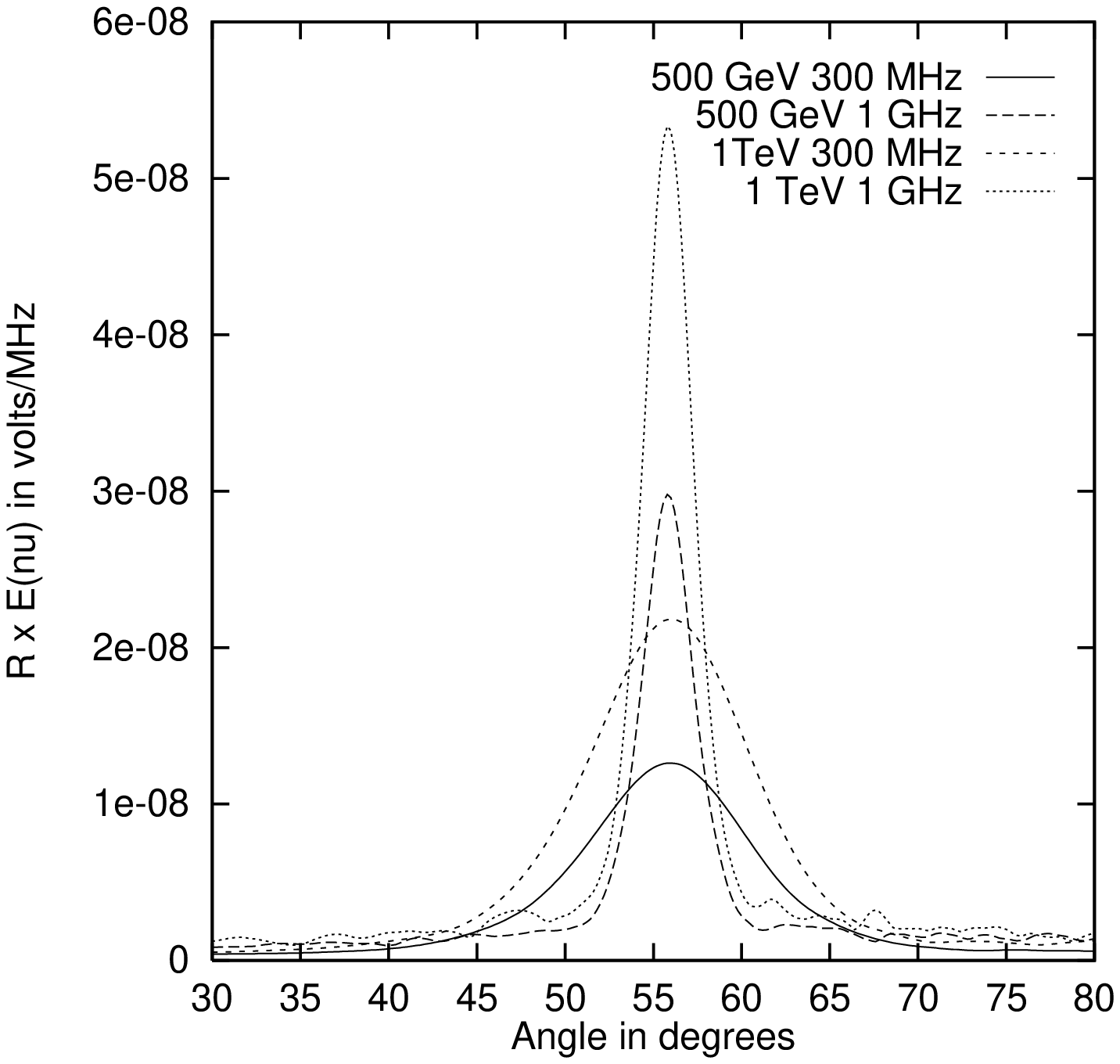,height=2.8in,width=2.8in}}
\vspace{10pt}
\caption{(a) $(dE/dx)_{ionization}$ calculated from GEANT output. The
average is over 500 electron tracks with 0.1 $MeV$ kinetic energy
threshold. We also show theoretical Bethe-Bloch curve. (b)
Electromagnetic pulses from cascades in ice generated by GEANT. Pulse
height scales linearly and pulse width scales inversely with primary
energy. Fraunhoffer limit has been taken to calculate field.}
\label{fig:pul}
\end{figure}
In Fig. \ref{fig:pul}b, we calculate pulses\footnote{we use different
normalization here for the field, which requiring that one multiply
(\ref{eq:pulse2}) and (\ref{eq:pulse3}) by $2\sqrt{2\pi}$} from
$1\;TeV$ and $500\;GeV$ showers at $1\;GHz$ and $300\;MHz$
frequencies. The pulse height at the Cherenkov peak ($\approx 55.8^o$)
scales with the energy of the shower, and the Gaussian half width is
inversely proportional to the frequency which is analogous to a slit
diffraction pattern.
\section{Conclusion}
We find distinctive, coherent, Cherenkov radio frequency emission
from simulated electromagnetic cascades in ice with energies in the
$100\;GeV$ to $1\;TeV$ range. The multi-purpose, detector simulation
package GEANT provides a suitable modelleing of electromagnetic shower
details for this purpose, with the flexibility to include hadronic
cascades in future studies. This work also serves as an independent
treatment of the problem, which has been studied previously
\cite{zhs92}. The GEANT generated showers are qualitatively in
agreement with those in \cite{zhs92}. The showers differ in several
details; however our far-field calculation of the pulse from a shower
gives the same result up to a $GHz$ frequency as that of \cite{zhs92}
when both are applied to the same shower simulation. This has been
checked explicitly by calculating the field using our field code and
the track informtions from ZHS code with the direct field output from
ZHS code for the same shower.  The differences between a GEANT and a
ZHS shower of initial energy $100\;GeV$ is summarized in Table II.
The origin of differences between the two Monte Carlos is under study.

Our direct calculation of Moliere radius, critical energy and shower
population as a function of shower depth using GEANT all agree within
a few percent with measurements or with published EGS4 simulations for
iron and lead \cite{pdg96}. The extension to ice is straightforward
and we expect the same level of reliability here.

Our results agree with the theoretically expected \cite{rossi} linear
scaling of number of particles at shower maximum and logarithimic
scaling of depth at shower maximum with energy. These are not
straightforward results, since the theoretical predictions are based
on inclusion of bremmstrahlung and pair production only, while the
physics at the shower maximum is not dominated by these processes. It
includes important contributions from Compton, Bhabha and Moller
processes as well, which give rise to charge imbalance and,
consequently, coherent radio Cherenkov field emission.

\begin{center} 
\begin{tabular}{lcc} 
\multicolumn{3}{c}{Table II Comparisons between GEANT and ZHS Monte
Carlos}\\ 
\multicolumn{3}{c}{(averaged over 20 showers each)}\\
\hline \hline
Quantity & GEANT &  ZHS \\
\hline
Primary Energy ($GeV$) & $100$ & $100$ \\
Total Energy Threshold ($MeV$) & $0.611$ & $0.611$ \\
Total absolute track length ($meter$) & $398.8 \pm 4.7$ & $642$ \\
Total projected ($e+p$) track length ($meter$) & $374.4 \pm 4.3$
& $518.7$ \\
({\it sum of electron and positron track lengths} & & \\
{\it projected along the shower axis}) & & \\
Total projected ($e-p$) track length ($meter$) & $70.0 \pm 8.4$
& $131.2$ \\
({\it difference of electron and positron track lengths} & & \\
{\it projected along the shower axis}) & & \\
Position of the shower max. (radiation length) & 7 & 7 \\
Number of particles ($e+p$) at shower max. & $111 \pm 26$
& $155 \pm 45$ \\
Excess electrons ($e-p$) at shower max. & $20 \pm 11$
& $37 \pm 14$ \\
Fractional charge excess at the shower max & $\sim 18\%$& $\sim 24\%$ 
\\
Cherenkov peak at $1\;GHz$ ($Volts/MHz$) & $7.46\times 10^{-9}$
& $1.08\times 10^{-8}$ \\
\hline \hline
\end{tabular}
\end{center}

{\bf Acknowledgement} Thanks to J. Alvarez-Mu\~niz, T. Bolton,
R. Buniy, G. Frichter, F. Halzen, J. Ralston, D. Seckel and E. Zas
for help and advice at various stages. This work is supported in part
by the NSF, the DOE, the University of Kansas General Research Fund,
the RESEARCH CORPORATION and the facilities of the Kansas Institute
for Theoretical and Computational Science.

\newcommand{\noopsort}[1]{} \newcommand{\printfirst}[2]{#1}
  \newcommand{\singleletter}[1]{#1} \newcommand{\switchargs}[2]{#2#1}

\end{document}